\documentclass[a4paper]{jpconf}

\usepackage[numbers,square,sort&compress]{natbib}
\usepackage{graphicx}
\usepackage{bm}

\begin{document}
\title{Inelastic neutron scattering study of phonon anomalies in
La$_{1.5}$Sr$_{0.5}$NiO$_{4}$}

\author{R Kajimoto$^1$, M Fujita$^2$, K Nakajima$^1$, K Ikeuchi$^3$, Y Inamura$^1$, M Nakamura$^1$ and T Imasato$^4$}

\address{$^1$ J-PARC Center, Japan Atomic Energy Agency, Tokai, Ibaraki
319-1195, Japan}

\address{$^2$ Institute for Materials Research, Tohoku University,
Sendai, Miyagi 980-8577, Japan}

\address{$^3$ Research Center for Neutron Science and Technology,
Comprehensive Research Organization for Science and Society, Tokai,
Ibaraki 319-1106, Japan}

\address{$^4$ Department of Physics, Tohoku University, Sendai, Miyagi
980-8578, Japan}

\ead{ryoichi.kajimoto@j-parc.jp}

\begin{abstract}
 The high-energy phonons in La$_{1.5}$Sr$_{0.5}$NiO$_{4}$, in which the
 checkerboard charge ordering is formed, was investigated by the
 inelastic neutron scattering. We found that the longitudinal
 modes show strong anomalies compared with La$_{2}$NiO$_{4}$. We argue
 the similarity and difference in the phonon anomalies between the
 present sample and the preceding works of different compositions.
\end{abstract}


La$_{2-x}$Sr$_{x}$NiO$_{4}$ is isostructural to one of the typical
cupurate superconductors La$_{2-x}$Sr$_{x}$CuO$_{4}$. When $x=0$, the
compound is an antiferromagnetic insulator, where $S=1$ spins on
Ni$^{2+}$ (3d$^8$) ions order antiferromagnetically. Similarly to the
cupurates, one hole per Ni ion is introduced in the NiO$_{2}$ planes by
the substitution of a La$^{3+}$ ion with a Sr$^{2+}$ ion. In the course
of exploring the origin of the superconductivity in the cupurates, the
role of phonons have been extensively studied, and it was revealed that
the hole doping induces a strong anomaly in the Cu-O bond stretching
mode~\cite{reznik10}. The phonon anomaly was regarded as a result of the
electron-lattice coupling associated with charge inhomogeneity in the
CuO$_{2}$ planes~\cite{reznik06}. In the nickelates, the
electron-lattice coupling is considered to be larger than the cupurates,
as the charge inhomogeneity appears in the more obvious form of static
charge stripes along the direction diagonal to the Ni square
lattice~\cite{chen93,tranquada94,tranquada96,yoshizawa00}. Then, phonons
in the nickelates have been studied by inelastic neutron scattering
(INS) to investigate whether the doped holes or charge stripes affect
particularly the bond-stretching modes, which are the highest-energy
longitudinal optical (LO) modes in this system. Pintschovius \textit{et
al.} reported that the LO modes in non-doped La$_{2}$NiO$_{4}$ show only
weak dispersion at $\sim$85~meV~\cite{pintschovius01}. However, they
found that the LO mode propagating along the [100] direction is
susceptible to the non-stoichiometry, and observed similar softening to
the cupurates in
La$_{1.9}$NiO$_{3.93}$~\cite{pintschovius01,pintschovius89}. A study on
powder samples of La$_{2-x}$Sr$_{x}$NiO$_{4}$ by McQueeney \textit{et
al.} showed that the $\sim$85~meV band in the phonon density of states
(DOS) splits into two sub-bands at $\sim$75~meV and $\sim$85~meV at $x
\ge 1/3$~\cite{mcqueeney98}. Tranquada \textit{et al.} reported that in
a single crystal of La$_{1.69}$Sr$_{0.31}$NiO$_{4}$, the LO mode along
[100] shows the similar softening, while that along [110] shows a
splitting of the same magnitude as the softening. However, the anomalies
in the phonons are independent on the charge stripe wave vector in this
compound, $\bm{q}_\mathrm{CO} = (0.31,0.31,0)$~\cite{tranquada02}.

In the present study, we performed an INS study of the high-energy
phonons in a single crystal of La$_{1.5}$Sr$_{0.5}$NiO$_{4}$. The wave
vector of the charge stripes is proportional to $x$, and in this
composition, the charge ordering results in a nearly checkerboard
ordering of Ni$^{2+}$ and Ni$^{3+}$ sites: The checkerboard charge
ordering with $\bm{q}_\mathrm{CO} = (1/2,1/2,0)$ is formed below
$\sim$480~K, and it is taken over by incommensurate charge order with
$\bm{q}_\mathrm{CO} \sim (0.44,0.44,0)$ below
$\sim$180~K~\cite{yoshizawa00,kajimoto03}. We found that the dispersions
of the phonon modes in $x=0.5$ show anomalies similar to those observed
in the $x=0.31$ compound. However, we also found some difference between
the two compositions.


Single crystals of La$_{1.5}$Sr$_{0.5}$NiO$_{4}$ were grown by the
floating-zone method. The crystal structure is tetragonal with the space
group $I4/mmm$~\cite{millburn99}. The lattice constants determined by
powder x-ray diffraction are $a = 3.814$~{\AA} and $c =
12.74$~{\AA}. Four crystal rods, each of which is $\sim$5~mm in diameter
and $\sim$30~mm in length, were assembled for the present work. The INS
measurement was performed on the chopper spectrometer 4SEASONS at
J-PARC~\cite{siki}. The incident energy was $E_\mathrm{i} = 111$~meV
with the energy resolution being 10~meV at the elastic scattering
condition.  We aligned the crystals so that the [001] axis is parallel
to the incident beam and the [110] axis is in the horizontal plane. Due
to this crystal orientation, the observed data may include the
contributions from out-of-plane phonon modes. However, they are expected
to be small, because the high-energy phonon modes observed in the
current experiment arise mainly from in-plane polarized oxygen
vibrations~\cite{mcqueeney98,pintschovius01}. We converted the raw data
taken at $\sim$5~K to a histogram of the intensity proportional to the
dynamical structure factor $S(\bm{Q},\hbar\omega)$, where $\bm{Q}$ and
$\hbar\omega$ are momentum and energy transfers, using the software
package Utsusemi~\cite{utsusemi}. In contrast to the previous INS
studies using triple-axis
spectrometers~\cite{pintschovius01,pintschovius89,tranquada02}, scanning
$Q_x = Ha^*$, $Q_y = Kb^*$, or $\hbar\omega$ simultaneously changes the
value of $Q_z = Lc^*$ in the present study using the chopper
spectrometer. Actually, for example, $(H,K) = (0,0)$, $(1,0)$, $(1,1)$
correspond to $(H,K,L) = (0,0,7.66)$, $(1,0,8.48)$, $(1,1,9.43)$ at
$\hbar\omega = 85$~meV, respectively. However, considering the layered
structure of the sample, we ignore $Q_z$ and express $\bm{Q}$ in terms
of $(H,K)$. Though this assumption is not strictly correct, it is
supported by the weak $Q_z$ dependence of the highest-energy modes
throughout the Brillouin zone observed in
La$_{1.9}$NiO$_{3.87}$~\cite{pintschovius89}. Since the obtained
intensity decreases as a function of $\hbar\omega$, we further divided
the intensity by $\hbar\omega$ for clarity in a high $\hbar\omega$
region. To correct the $\hbar\omega$-dependent background, we estimated
it by fitting the background of the intensity in the region of
$0.9\!<\!H\!<\!1.1$ and $-0.1\!<\!K\!<\!0.1$ to a quadratic function of
$\hbar\omega$, then subtracted it from the data.


\begin{figure}[tb]
 \includegraphics[width=0.6\hsize]{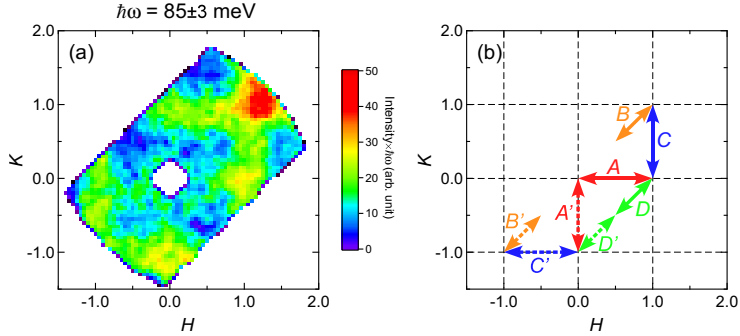}\hspace{0.05\hsize}%
 \begin{minipage}[b]{0.35\hsize}
  \caption{\label{Qmap}(a) Intensity map of the excitations in
  La$_{1.5}$Sr$_{0.5}$NiO$_{4}$ at $\sim$5~K, which is cut at
  $\hbar\omega = 85\pm3$~meV and displayed on the $H$-$K$ plane. The
  intensity is smoothed by a Gaussian filter. (b)
  Directions in the $H$-$K$ plane along which the data in
  figure~\ref{QEmap} are shown (see caption of figure~\ref{QEmap} and
  text).}
 \end{minipage}
\end{figure}

First, to survey the overall structure of the high-energy phonons, we
investigated a $Q$ map of the neutron scattering intensity.  Figure
\ref{Qmap}(a) shows the intensity map at $\hbar\omega = 85$~meV on the
$H$-$K$ plane. This energy is almost equal to that of the
bond-stretching mode in La$_{2}$NiO$_{4}$~\cite{pintschovius01}. A
characteristic intensity modulation is observed in Fig.\ \ref{Qmap},
which reflects some dispersions of the high-energy modes. There are
strong spots at $(H,K) = (1,1)$, $(\pm 1,0)$, and $(0, \pm 1)$. In
addition, the intensity shows streaks along the $(H,\pm 1)$ and $(\pm
1,K)$ lines. These spots and streaks suggest that the phonon mode along
these lines shows a weak dispersion, while those along the other
directions show stronger dispersions.

\begin{figure}[tb]
 \includegraphics[width=0.55\hsize]{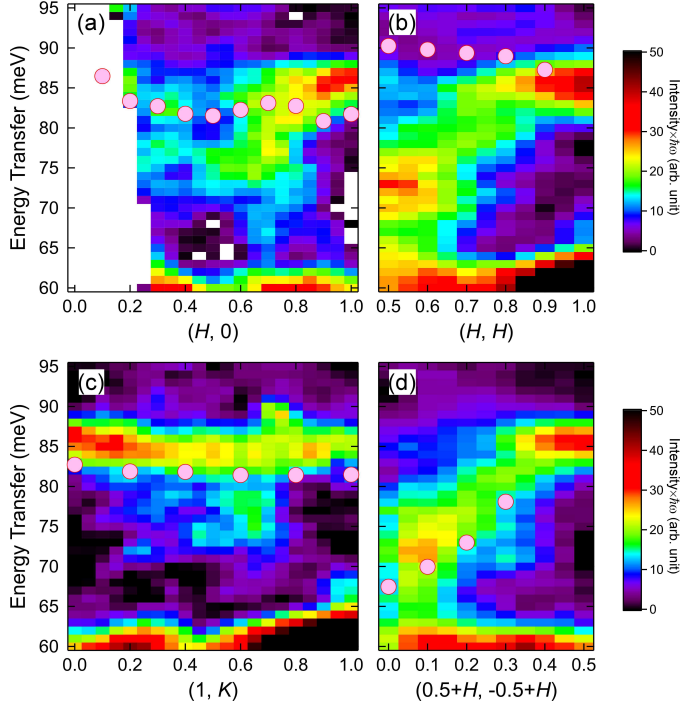}\hspace{0.05\hsize}%
 \begin{minipage}[b]{0.4\hsize}
 \caption{\label{QEmap}Intensity map of the high-energy excitations in
  La$_{1.5}$Sr$_{0.5}$NiO$_{4}$ at $\sim$5~K on the $Q$-$\hbar\omega$
  planes. The intensity is smoothed by a Gaussian filter. Each data is
  sliced along each $Q$ direction with a width of $H$ or $K = \pm 0.1$.
  The directions of the $Q$ axis of (a), (b), (c), and (d) are $A+A'$,
  $B+B'$, $C+C'$, and $D+D'$ in figure~\ref{Qmap}(b), respectively (see
  text). Circles indicate the energies of phonons observed in
  La$_{2}$NiO$_{4}$~\cite{pintschovius01}.}
 \end{minipage}
\end{figure}

Next, to investigate the $\bm{Q}$ dependence of the high-energy phonons
in more detail, we cut the data along the lines shown in figure
\ref{Qmap}(b). Figure \ref{QEmap} shows thus obtained intensity maps as
functions of $Q$ and $\hbar\omega$. Here, data in (a), (b), (c), and (d)
correspond to the cuts along the lines indicated by $A$, $B$, $C$, and
$D$ in figure \ref{Qmap}(b), respectively. To improve the statistics, we
averaged the symmetrically identical data along $A$ and $A'$, $B$ and
$B'$, $C$ and $C'$, and $D$ and $D'$. Clear excitations are observed in
the energy region between $\sim$70~meV and $\sim$90~meV where the
in-plane modes of phonons were observed in the previous studies of
nickelates and cupurates with the similar crystal
structures~\cite{reznik10,reznik06,pintschovius01,pintschovius89,mcqueeney98,tranquada02},
which should be an evidence of the observed excitations are caused by
the phonons.  Figures \ref{QEmap}(a) and \ref{QEmap}(b) reflect
scattering from the longitudinal optical (LO) modes along [100] and
[110], while figures \ref{QEmap}(c) and \ref{QEmap}(d) reflect that from
the transverse optical (TO) modes along [100] and [110],
respectively. We also show the energies of phonons observed in
La$_{2}$NiO$_{4}$ along the equivalent directions
(circles)~\cite{pintschovius01}.

The energies of the LO modes in La$_{2}$NiO$_{4}$ show weak $Q$
dependences both along [100] and [110], as shown by circles in figures
\ref{QEmap}(a) and \ref{QEmap}(b). In contrast, the scattering spectra
of La$_{1.5}$Sr$_{0.5}$NiO$_{4}$ in figures \ref{QEmap}(a) and
\ref{QEmap}(b) show quite different dispersions from
La$_{2}$NiO$_{4}$. They show large $Q$ dependences with the minimum
energies of $\sim$73~meV at $H=0.5$ along both the directions. We should
note that this minimum energy is almost equal to that of the sub-band in
the phonon DOS appearing at $x>1/3$~\cite{mcqueeney98}, which suggests
these anomalies in the LO modes are the origin of the $\sim$75~meV
sub-band. On the other hand, for the TO modes in figures \ref{QEmap}(c)
and \ref{QEmap}(d), the observed dispersions are almost similar to
La$_{2}$NiO$_{4}$, though the energies are slightly higher. The hole
doping and charge ordering do not affect the energies of the TO modes
except for slight hardenings.

The observed anomalies in the LO modes as well as their minimum energies
at $\sim$73~meV are apparently similar to those observed in
$x=0.31$~\cite{tranquada02}. In $x=0.31$, though the energy of the LO
phonon along [100] continuously softens as a function of $Q$, that along
[110] shows a splitting into two modes. Moreover, the latter anomaly is
independent on $\bm{q}_\mathrm{CO}$. Tranquada \textit{et al.}
interpreted the splitting of the LO mode is caused by the local
breathing motion of O ions about the hole-doped Ni sites (Ni$^{3+}$ sites). The
similarities between the phonon anomalies in $x=0.5$ and those in
$x=0.31$ supports their idea that the doped holes cause local effects on
the bond-stretching phonons independent on the hole concentration.

However, there is some difference between the two compounds. Contrary to
$x=0.31$, in $x=0.5$, the dispersion of the LO mode along [110] shows a
clear softening with its minimum at (0.5,0.5) [figure \ref{QEmap}(b)],
though there is very weak $Q$ independent intensity at $\sim$85~meV. In
other words, the softening behavior is superior to the splitting behavior
along [110] in this compound. On the other hand, the splitting behavior
is more clearly observed in the data along [100]
[figure~\ref{QEmap}(a)]. The difference between the present $x=0.5$
sample and $x=0.31$ may be related to the difference in types of the
charge ordering. In the case of the checkerboard charge ordering in
$x=0.5$, the holes enter every two Ni sites, and Ni$^{2+}$ sites and
Ni$^{3+}$ sites always share the in-plane O ions. Therefore, the
vibrations of the shared in-plane O ions may become more coherent along
the direction of the charge-ordering wave vector, [110], resulting in
the softening of the dispersion.


In conclusion, we performed an INS study on
La$_{1.5}$Sr$_{0.5}$NiO$_{4}$ to investigate the effects of hole doping
on the high-energy phonon modes. We found that the longitudinal modes
show clearly different dispersions compared with La$_{2}$NiO$_{4}$,
though the transverse modes are quite similar to those in
La$_{2}$NiO$_{4}$. In particular, the longitudinal mode along the
direction diagonal to the Ni square lattice shows a clear softening, in
contrast to $x=0.31$ where the mode shows a
splitting~\cite{tranquada02}. We interpret the difference in the phonon
anomalies between $x=0.5$ and $x=0.31$ as the difference in the type of
charge ordering.

\ack

The experiment on 4SEASONS was conducted under the project numbers
2012I0100 and 2012I0101.


\bibliography{LPBMS_LSNO_131013_resub}

\end{document}